\documentclass[12pt]{iopart}

\newtheorem{pro}{Proposition}

\usepackage{iopams}
\begin{document}

\title[On some integrable lattice related  to the Itoh-Narita-Bogoyavlenskii lattice]{On some integrable lattice related by the Miura-type transformation to the Itoh-Narita-Bogoyavlenskii lattice}

\author{Andrei K Svinin}

\address{Institute for System
Dynamics and Control Theory, Siberian Branch of
Russian Academy of Sciences, Russia}
\ead{svinin@icc.ru}
\begin{abstract}
We show that by  Miura-type transformation the Itoh-Narita-Bogoyavlenskii lattice, for any $n\geq 1$, is related to some differential-difference (modified) equation. We present corresponding integrable hierarchies in its explicit form. We study the elementary Darboux transformation for modified equations.

\end{abstract}

\vspace{2pc}


\submitto{J. Phys. A: Math. Theor.}

\section{Introduction}

The equation
\begin{equation}
v_i^{\prime}=-\frac{1}{v_{i-1}-v_{i+1}}
\label{vi}
\end{equation}
is known to be one of a representative in the class of integrable differential-difference equations of Volterra-type form $v_i^{\prime}=f(v_{i-1}, v_i, v_{i+1})$ \cite{Yamilov}. As is known, equation (\ref{vi}) is related to the Volterra lattice $a_i^{\prime}=a_i(a_{i-1}-a_{i+1})$ by the Miura-type transformation \cite{Levi}
\begin{equation}
a_i=\frac{1}{(v_{i-2}-v_i)(v_{i-1}-v_{i+1})}.
\label{Miura}
\end{equation}
The Volterra lattice gives natural integrable discretization of the Korteweg-de Vries (KdV) equation.  A more general class of integrable differential-difference equations sharing this property is given by the Itoh-Narita-Bogoyavlenskii (INB) lattice \cite{Bogoyavlenskii, Itoh, Narita}
\begin{equation}
a_i^{\prime}=a_i\left(\sum_{j=1}^na_{i-j}-\sum_{j=1}^na_{i+j}\right).
\label{INB}
\end{equation}
This equation primarily known in the literature as the Bogoyavlenskii lattice was considered in \cite{Narita} as natural integrable generalization of the Volterra lattice --- the equation admitting bilinear Hirota's form and  soliton solutions. In \cite{Itoh}, Itoh used Lotka-Volterra finite-dimensional dynamical systems, which can be obtained by imposing on (\ref{INB}) specific periodicity conditions $a_{i+2n+1}=a_i$.

The Volterra lattice and its generalization (\ref{INB}) share the property that it admit integrable hierarchy. This means that the flow defined on suitable phase space by the evolution equation (\ref{INB}) can be included into infinite set of the pairwise commuting flows. We have shown in \cite{Svinin1, Svinin2} that corresponding integrable hierarchies for (\ref{INB}) and for some number of integrable lattices are directly related to the KP hierarchy.

The goal of this paper is to present some class of differential-difference equations each of which is related to the INB lattice with any $n\geq 1$ by some Miura-type transformation generalizing (\ref{Miura}). In \cite{Svinin4, Svinin5}, we showed explicit form for INB lattice hierarchy in terms of some homogeneous polynomials $S_s^l[a]$. Using these results, we give the explicit form of modified evolution equations on the field $v=v_i$ governing corresponding  integrable hierarchies in terms of some rational functions $\tilde{S}_s^l[v]$. In section \ref{sec:3}, we study Darboux transformation for underlying linear equations of modified integrable hierarchies and as a result derive some discrete quadratic equation parameterized by $n\geq 1$, which is shown to relate two solutions $\{v_i\}$ and $\{\bar{v}_i\}$ of modified equation yielding the first flow in corresponding integrable hierarchy. We claim but do not prove in this paper that this algebraic relation is also compatible  with higher flows of modified integrable hierarchy. We can interpret the quadratic equation $F[v, \bar{v}]=0$ as a two-dimensional equation on $\{v_{i, j}\}$. We show that this  equation can be written in the form of the relation $I_{i+1, j}=I_{i, j}$ with an appropriate integral $I_j$.  In the case $n=1$, this two-dimensional equation can be treated as a deautonomization of the lattice potential KdV (lpKdV) equation $I_{i, j}=\left(v_{i, j}-v_{i+1, j+1}\right)\left(v_{i+1, j}-v_{i, j+1}\right)=c$. A similar relation for any $n\geq 2$ gives some generalization of the lpKdV equation.

\section{Miura-type transformation for INB lattice}

\subsection{The INB lattice and its hierarchy}

Let us consider the INB lattice (\ref{INB}) together with auxiliary linear equations on the KP wavefunction:
\begin{equation}
z\psi_{i+n}+a_{i+n}\psi_i=z\psi_{i+n+1},\;\;\;
\psi_i^{\prime}=z\psi_{i+n}-\left(\sum_{j=1}^na_{i+j-1}\right)\psi_i.
\label{linear}
\end{equation}
As was shown in \cite{Svinin4, Svinin5}, the linear equation governing higher flows of equation (\ref{INB}) can be written as
\begin{equation}
\partial_s\psi_i=z^s\psi_{i+sn}+\sum_{j=1}^sz^{s-j}(-1)^jS^j_{sn-1}(i-(j-1)n)\psi_{i+(s-j)n}.
\label{int1}
\end{equation}
Coefficients of this equation are defined by some special polynomials $S^l_s[a]$. More exactly, let
\[
S^l_s(y_1,\ldots, y_{s+(l-1)n+1})=\sum_{1\leq\lambda_l\leq\cdots\leq\lambda_1\leq s+1}\left(\prod_{j=1}^ly_{\lambda_j+n(j-1)}\right).
\]
Then $S^l_s(i)\equiv S^l_s(a_i,\ldots, a_{i+s+(l-1)n})$. Integrable hierarchy of the flows for the INB lattice (\ref{INB}) is defined by the totality of evolution differential-difference equations \cite{Svinin4, Svinin5}
\begin{equation}
\partial_sa_i=(-1)^sa_i\left\{S_{sn-1}^s(i-(s-1)n+1)-S_{sn-1}^s(i-sn)\right\}
\label{higher}
\end{equation}
for $s\geq 1$.

\subsection{Miura-type transformation}

It is known that the noninvertible substitution
\[
a_i=\prod_{k=1}^{n+1}u_{i+k-1}
\]
serves as the Miura-type transformation between the INB lattice and its modified version \cite{Bogoyavlenskii}
\[
u_i^{\prime}=u_i^2\left(\prod_{k=1}^nu_{i-k}-\prod_{k=1}^nu_{i+k}\right).
\] 
This equation can be written in the form of the following conservation law:
\[
\left(\frac{1}{u_i}\right)^{\prime}=V_{i-2n}-V_{i-n+1}\;\;\;\mbox{with}\;\;\; V_i=-\prod_{k=n}^{2n-1}u_{i+k}.
\]
The latter suggests to introduce the potential $1/u_i=v_{i-2n}-v_{i-n+1}$. Therefore, we come to the rational substitution
\begin{equation}
a_i=\prod_{j=n}^{2n}\frac{1}{v_{i-j}-v_{i-j+n+1}}
\label{av}
\end{equation}
which is a generalization of (\ref{Miura}). It sends solutions of the equation
\begin{equation}
v_i^{\prime}=-\prod_{j=1}^n\frac{1}{v_{i-j}-v_{i-j+n+1}}
\label{singular}
\end{equation}
to solutions of the INB equation (\ref{INB}). Moreover, we have the following.

\begin{pro} \label{p}
Integrable hierarchy of lattice (\ref{singular}) is given by evolution differential-difference equations\footnote{It is supposed, in (\ref{int}), that $\tilde{S}^l_s[v]\equiv S^l_s[a]$, where the field $a$ is expressed via $v$ in virtue of (\ref{av}).}
\begin{equation}
\partial_sv_i=(-1)^s\tilde{S}_{sn-1}^{s-1}(i-(s-2)n)\prod_{j=1}^n\frac{1}{v_{i-j}-v_{i-j+n+1}}.
\label{int}
\end{equation}
\end{pro}

Therefore, we obtained integrable hierarchy (\ref{int}) related to the INB lattice hierarchy via substitution (\ref{av}). 
Since there exist a number of Miura-type transformations, which relate the INB lattice to its modifications (see, for example \cite{Bogoyavlenskii, Papageorgiou}) the question arises how many integrable equations related by the Miura-type transformations do exist?  

\subsection{Auxiliary linear equations}

Introduce the function $\gamma_i$,  such that
\begin{equation}
\frac{\gamma_{i+1}}{\gamma_i}=\frac{1}{v_{i-n}-v_{i+1}}.
\label{gamma}
\end{equation}
Clearly, in terms of $\gamma_i$
\[
a_i=\prod_{j=1}^{n+1}\frac{\gamma_{i-n+j}}{\gamma_{i-n+j-1}}.
\]
\begin{pro} \label{l:1}
In virtue of (\ref{int}) and (\ref{gamma}),
\begin{equation}
\partial_s\gamma_i=(-1)^s\tilde{S}^s_{sn-1}(i-(s-1)n)\gamma_i.
\label{lemma}
\end{equation}
\end{pro}

 Let $\psi_i=\gamma_i\phi_i$. In terms of this new wavefunction, linear equations (\ref{linear}) become
\begin{equation}
z(v_i-v_{i+n+1})\phi_{i+n}+\phi_i=z\phi_{i+n+1},
\label{linear1}
\end{equation}
\begin{equation}
\phi_i^{\prime}=z\phi_{i+n}\prod_{j=1}^n\frac{1}{v_{i-j}-v_{i-j+n+1}}.
\label{linear2}
\end{equation}
To derive these equations, we take into account proposition \ref{l:1}, namely, that
\[
\gamma_i^{\prime}=-\tilde{S}^1_{n-1}(i)\gamma_i=-\sum_{j=1}^n\left(\prod_{k=n}^{2n}\frac{1}{v_{i-k+j-1}-v_{i-k+j+n}}\right)\gamma_i.
\]
In a similar way we get the linear equation
\begin{equation}
\partial_s\phi_i=z^s\frac{\gamma_{i+sn}}{\gamma_i}\phi_{i+sn}+\sum_{j=1}^{s-1}(-1)^jz^{s-j}\frac{\gamma_{i+(s-j)n}}{\gamma_i}\tilde{S}^j_{sn-1}(i-(j-1)n)\phi_{i+(s-j)n}
\label{linear4}
\end{equation}
governing all the flows (\ref{int}).

\section{Darboux transformation for linear equations (\ref{linear1}) and (\ref{linear2})}
\label{sec:3}

\subsection{Darboux transformation}

Let us show that linear equations (\ref{linear1}) and (\ref{linear2}) admit Darboux transformation
\begin{equation}
\bar{\phi}_i=\phi_{i+1}+(v_{i+1}-\bar{v}_i)\phi_i.
\label{Darboux}
\end{equation}
It is worth remarking that the latter does not depend on $n$. It is simple exercise to verify that (\ref{Darboux}) yields Darboux transformation for linear equation (\ref{linear1}) if the relation
\begin{equation}
(v_i-v_{i+n+1})(v_{i+1}-\bar{v}_i)=(v_{i+n+1}-\bar{v}_{i+n})(\bar{v}_i-\bar{v}_{i+n+1})
\label{equation-1}
\end{equation}
holds. One can rewrite this equation as
\begin{equation}
(v_i-v_{i+n+1})(v_{i+1}-\bar{v}_{i+n+1})=
(v_i-\bar{v}_{i+n})(\bar{v}_i-\bar{v}_{i+n+1})
\label{equation0}
\end{equation}
and
\begin{equation}
(v_i-\bar{v}_{i+n})(v_{i+1}-\bar{v}_i)=
(v_{i+1}-\bar{v}_{i+n+1})(v_{i+n+1}-\bar{v}_{i+n})
\label{equation1}
\end{equation}
and
\begin{equation}
v_iv_{i+1}-v_{i+1}v_{i+n+1}+v_{i+n+1}\bar{v}_{i+n+1}-v_i\bar{v}_i+\bar{v}_i\bar{v}_{i+n}-\bar{v}_{i+n}\bar{v}_{i+n+1}=0.
\label{equation}
\end{equation}
We have the following.
\begin{pro}
\label{pro:3}
Relation (\ref{equation}) is compatible with the flow given by (\ref{singular}).
\end{pro}

It is natural to expect that  relation (\ref{equation}) in fact is compatible with all the higher flows (\ref{int}), but this question we leave for subsequent studies. Taking into account proposition (\ref{pro:3}), one can check that the relation (\ref{Darboux}) yields elementary Darboux transformation for  linear equation (\ref{linear2}).

\subsection{Generalized lpKdV equation}

We observe that
\begin{equation}
I_i=\left(v_i-\bar{v}_{i+n}\right)\prod_{k=1}^n\left(v_{i+k}-\bar{v}_{i+k-1}\right)
\label{integral}
\end{equation}
is an integral for discrete equation (\ref{equation}).  Indeed, making use of (\ref{equation1}), one can easily check that $I_{i+1}=I_i$. Identifying $v_i=v_{i, j}$, $\bar{v}_i=v_{i, j+1}$ and $I_i=I_{i, j}$, we can interpret the relation (\ref{equation}) as a two-dimensional discrete equation with integral (\ref{integral}). In the case $n=1$,  equation (\ref{equation}) and its integral (\ref{integral}) are specified as
\begin{eqnarray}
v_{i, j}v_{i+1, j}&-&v_{i+1, j}v_{i+2, j}+v_{i+2, j}v_{i+2, j+1} \nonumber \\
                  &-&v_{i, j}v_{i, j+1}+v_{i, j+1}v_{i+1, j+1}-v_{i+1, j+1}v_{i+2, j+1}=0
\label{equation2}
\end{eqnarray}
and
\[
I_{i, j}=\left(v_{i, j}-v_{i+1, j+1}\right)\left(v_{i+1, j}-v_{i, j+1}\right),
\]
respectively. Assigning some values: $I_{i, j}=c_j$, we are led to the nonautonomous equation
\[
\left(v_{i, j}-v_{i+1, j+1}\right)\left(v_{i+1, j}-v_{i, j+1}\right)=c_j
\]
which is obviously equivalent to (\ref{equation2}). In particular case when $c_j=c$ with some constant $c$, we come to the  lpKdV equation \cite{Nijhoff1, Nijhoff2}
\begin{equation}
\left(v_{i, j}-v_{i+1, j+1}\right)\left(v_{i+1, j}-v_{i, j+1}\right)=c
\label{equation3}
\end{equation}
being the potential version of Hirota's discrete KdV equation \cite{Hirota}. It is also known as the $H_1$ equation in Adler-Bobenko-Suris classification \cite{Adler}. Recall that the relation (\ref{equation3}) with more general right-hand side $c_j-b_i$ where $\{c_j, b_i\}$ are some set of parameters of Darboux transformations, naturally appear as a nonlinear superposition principle for the potential KdV equation \cite{Wahlquist}. This fully deautonomized version of lpKdV equation is also referred to as the lpKdV equation.

Therefore, equation (\ref{equation2}) can be treated as the (partial) deautonomization of (\ref{equation3}). In the general case, we have the following equation:
\begin{equation}
\left(v_{i, j}-v_{i+n, j+1}\right)\prod_{k=1}^n\left(v_{i+k, j}-v_{i+k-1, j+1}\right)=c,
\label{equation4}
\end{equation}
being in a sense a generalization of the lpKdV equation (\ref{equation3}). As can be checked, the  deautonomization of this equation given by a change $c\mapsto c_j$ yields the equation 
\[
v_{i, j}v_{i+1, j}-v_{i+1, j}v_{i+n+1, j}+v_{i+n+1, j}v_{i+n+1, j+1}
\]
\[
-v_iv_{i, j+1}+v_{i, j+1}v_{i+n, j+1}-v_{i+n, j+1}v_{i+n+1, j+1}=0.
\]

\section{Conclusion}

We have shown in this paper an infinite number of integrable hierarchies (\ref{int}) related via the noninvertible substitution to INB lattice hierarchies given by evolution equations (\ref{higher}). They appear also as compatibility conditions of  linear equations (\ref{linear1}) and (\ref{linear4}). In section \ref{sec:3}, we studied Darboux transformation for linear equations. We have shown that the discrete equation (\ref{equation}) yields the elementary Darboux transformation.  In the case $n=1$, it can be obtained by a sort of deautonomization of the lpKdV equation (\ref{equation3}). The same is true in the general case. Namely, one can derive  equation (\ref{equation}) as the deautonomization of the generalized  lpKdV equation (\ref{equation4}).

\section{Acknowledgements}

We thank the referees for carefully reading the manuscript and for a number of remarks that enabled the presentation of the paper to be improved.


\appendix

\section{Proof of proposition \ref{p}}

We must to show that in virtue of equation (\ref{int}) the equation  (\ref{int1}) holds. We have
\begin{eqnarray}
\fl
\partial_sa_i&=&\partial_s\left(\prod_{j=n}^{2n}\frac{1}{v_{i-j}-v_{i-j+n+1}}\right)=\prod_{j=n}^{2n}\frac{1}{v_{i-j}-v_{i-j+n+1}}
\cdot\sum_{j=n}^{2n}\frac{\partial_sv_{i-j+n+1}-\partial_sv_{i-j}}{v_{i-j}-v_{i-j+n+1}}
\nonumber \\
\fl
&=&(-1)^s\prod_{j=n}^{2n}\frac{1}{v_{i-j}-v_{i-j+n+1}}\left\{\sum_{j=n}^{2n}\tilde{S}_{sn-1}^{s-1}(i-j-(s-3)n+1)\prod_{k=0}^n\frac{1}{v_{i-j-k+n}-v_{i-j-k+2n+1}}\right.
\nonumber \\
\fl
&&\left.-\sum_{j=n}^{2n}\tilde{S}_{sn-1}^{s-1}(i-j-(s-2)n)\prod_{k=0}^n\frac{1}{v_{i-j-k}-v_{i-j-k+n+1}} \right\}
\nonumber \\
\fl
&=&(-1)^sa_i\left\{\sum_{j=n}^{2n}a_{i+2n-j}S_{sn-1}^{s-1}(i-j-(s-3)n+1) \right.
\nonumber \\
\fl
&&\left.-\sum_{j=n}^{2n}a_{i+n-j}S_{sn-1}^{s-1}(i-j-(s-2)n)\right\}.
\nonumber
\end{eqnarray}
To proceed, we need to use the identity \cite{Svinin4,Svinin5}
\begin{equation}
S^l_s(i+1)-S^l_s(i)=a_{i+s+1}S^{l-1}_s(i+n+1)-a_{i+(l-1)n}S^{l-1}_s(i).
\label{identity}
\end{equation}
Taking the latter into account we obtain
\begin{eqnarray}
\partial_sa_i&=&(-1)^sa_i\sum_{j=n}^{2n}\left(S_{sn-1}^s(i-j-(s-2)n+1)-S_{sn-1}^s(i-j-(s-2)n)\right)
\nonumber \\
&=&(-1)^sa_i\left\{S_{sn-1}^s(i-(s-1)n+1)-S_{sn-1}^s(i-sn)\right\}.
\nonumber
\end{eqnarray}

\section{Proof of proposition \ref{l:1}}

On the one hand
\begin{equation}
\partial_s\left(\frac{\gamma_{i+1}}{\gamma_i}\right)=\frac{\gamma_{i+1}}{\gamma_i}\left(\frac{\partial_s\gamma_{i+1}}{\gamma_{i+1}}-\frac{\partial_s\gamma_i}{\gamma_i}\right).
\label{11}
\end{equation}
On the other hand
\begin{eqnarray}
\fl
\partial_s\left(\frac{\gamma_{i+1}}{\gamma_i}\right)&=&\partial_s\left(\frac{1}{v_{i-n}-v_{i+1}}\right)=\frac{\left(\partial_sv_{i+1}-\partial_sv_{i-n}\right)}{(v_{i-n}-v_{i+1})^2}
\nonumber \\
\fl
&=&(-1)^s\frac{1}{v_{i-n}-v_{i+1}}\left\{
\tilde{S}_{sn-1}^{s-1}(i-(s-2)n+1)\prod_{j=0}^n\frac{1}{v_{i-j}-v_{i-j+n+1}} \right.
\nonumber \\
\fl
&&\left.-\tilde{S}_{sn-1}^{s-1}(i-(s-1)n)\prod_{j=0}^n\frac{1}{v_{i-j-n}-v_{i-j+1}}
\right\}
\nonumber \\
\fl
&=&(-1)^s\frac{\gamma_{i+1}}{\gamma_i}\left\{a_{i+n}\tilde{S}_{sn-1}^{s-1}(i-(s-2)n+1)-a_i\tilde{S}_{sn-1}^{s-1}(i-(s-1)n)\right\}
\nonumber \\
\fl
&=&(-1)^s\frac{\gamma_{i+1}}{\gamma_i}\left\{\tilde{S}_{sn-1}^s(i-(s-1)n+1)-\tilde{S}_{sn-1}^s(i-(s-1)n)\right\}.
\label{22}
\end{eqnarray}
Remark that we used here the identity (\ref{identity}). Comparing (\ref{22}) with (\ref{11}), we come to (\ref{lemma}).

\section{Proof of proposition \ref{pro:3}}

Differentiating (\ref{equation}) in virtue of (\ref{singular}) yields
\[
\fl
\left(\bar{v}_i-v_{i+1}\right)\prod_{j=1}^n\frac{1}{v_{i-j}-v_{i-j+n+1}}+
\left(v_{i+n+1}-v_i\right)\prod_{j=0}^{n-1}\frac{1}{v_{i-j}-v_{i-j+n+1}}
\]
\[
\fl
+\left(v_{i+1}-\bar{v}_{i+n+1}\right)\prod_{j=-n}^{-1}\frac{1}{v_{i-j}-v_{i-j+n+1}}+
\left(\bar{v}_{i+n}-v_{i+n+1}\right)\prod_{j=-n}^{-1}\frac{1}{\bar{v}_{i-j}-\bar{v}_{i-j+n+1}}
\]
\begin{equation}
\fl
+\left(\bar{v}_{i+n+1}-\bar{v}_i\right)\prod_{j=-n+1}^{0}\frac{1}{\bar{v}_{i-j}-\bar{v}_{i-j+n+1}}+
\left(v_i-\bar{v}_{i+n}\right)\prod_{j=1}^n\frac{1}{\bar{v}_{i-j}-\bar{v}_{i-j+n+1}}=0.
\label{app}
\end{equation}
Thus, we must to prove that relation (\ref{app}) is the identity in virtue of (\ref{equation}). We are in position to prove that in fact two relations
\[
\left(\bar{v}_i-v_{i+1}\right)\prod_{j=1}^n\frac{1}{v_{i-j}-v_{i-j+n+1}}+
\left(v_{i+n+1}-v_i\right)\prod_{j=0}^{n-1}\frac{1}{v_{i-j}-v_{i-j+n+1}}
\]
\begin{equation}
+\left(v_i-\bar{v}_{i+n}\right)\prod_{j=1}^n\frac{1}{\bar{v}_{i-j}-\bar{v}_{i-j+n+1}}=0
\label{app1}
\end{equation}
and
\[
\left(v_{i+1}-\bar{v}_{i+n+1}\right)\prod_{j=-n}^{-1}\frac{1}{v_{i-j}-v_{i-j+n+1}}+
\left(\bar{v}_{i+n}-v_{i+n+1}\right)\prod_{j=-n}^{-1}\frac{1}{\bar{v}_{i-j}-\bar{v}_{i-j+n+1}}
\]
\begin{equation}
+\left(\bar{v}_{i+n+1}-\bar{v}_i\right)\prod_{j=-n+1}^{0}\frac{1}{\bar{v}_{i-j}-\bar{v}_{i-j+n+1}}=0
\label{app2}
\end{equation}
are both the identities in virtue of (\ref{equation}). Let us prove (\ref{app1}). Remark that one can write the second member in (\ref{app1}) as
\[
G_2=
\left(\bar{v}_{i+1}-\bar{v}_{i-n}\right)
\left(\prod_{j=1}^{n-1}\frac{1}{v_{i-j}-v_{i-j+n+1}}\right)\frac{1}{\bar{v}_{i-n}-\bar{v}_{i+1}}.
\]
In turn, using step-by-step relation (\ref{equation0}), one can rewrite the third member in (\ref{app1}) in the form
\[
G_3=
\left(v_{i-n+1}-\bar{v}_{i+1}\right)
\left(\prod_{j=1}^{n-1}\frac{1}{v_{i-j}-v_{i-j+n+1}}\right)\frac{1}{\bar{v}_{i-n}-\bar{v}_{i+1}}.
\]
Summing these two terms and using (\ref{equation-1}) we obtain
\begin{eqnarray}
G_2+G_3&=&\left(\prod_{j=1}^{n-1}\frac{1}{v_{i-j}-v_{i-j+n+1}}\right)\frac{v_{i-n+1}-\bar{v}_{i-n}}{\bar{v}_{i-n}-\bar{v}_{i+1}} \nonumber \\
        &=&\left(\prod_{j=1}^{n-1}\frac{1}{v_{i-j}-v_{i-j+n+1}}\right)\frac{v_{i+1}-\bar{v}_i}{v_{i-n}-v_{i+1}} \nonumber \\
        &=&\left(v_{i+1}-\bar{v}_i\right)\left(\prod_{j=1}^n\frac{1}{v_{i-j}-v_{i-j+n+1}}\right)=-G_1. \nonumber
\end{eqnarray}
Therefore we proved (\ref{app1}). The similar reasoning is needed to prove (\ref{app2}) and as a result (\ref{app}).



\end{document}